# Design of battery materials via defects and doping

**Khang Hoang**[*]

Center for Computationally Assisted Science and Technology & Department of Physics
North Dakota State University, Fargo, ND 58108, United States

## Abstract

This chapter illustrates the use of defect physics as a conceptual and theoretical framework for understanding and designing battery materials. It starts with a methodology for first-principles studies of defects in complex transition-metal oxides. The chapter then considers defects that are activated in a cathode material during synthesis, during measurements, and during battery use. Through these cases, it discusses possible defect landscapes in the material and their implications, guidelines for materials design via defect-controlled synthesis, mechanisms for electronic and ionic conduction and for electrochemical extraction and (re-)insertion, and effects of doping. Although specific examples are taken from studies of battery cathode materials, the computational approach and discussions are general and applicable to any ionic, electronic, or mixed ionic–electronic conducting materials.

## 1 Introduction

Crystallographic point defects are ubiquitous in solids. They can be electronic (e.g., electron and hole polarons) or ionic (vacancies, interstitials, and antisites); and intrinsic (i.e., native to a material) or extrinsic (substitutional or interstitial impurities incorporated into a material intentionally, i.e., as dopants, or unintentionally present). They can stay isolated or agglomerate and form complexes. Native defects, impurities, and defect complexes are often referred commonly to as *defects*. Defects can form during materials synthesis, during measurements, and/or during use, and be activated thermally, optically, electrochemically, or by other means. They can be neutral or, especially in semiconductors or insulators, positively or negatively charged.

As battery electrodes are often made of structurally and chemically complex materials, e.g., transition-metal oxides and polyanionic compounds, they are even more prone to defects. The presence of some of these defects can be vital or detrimental to the battery performance. Examples of good defects include small hole and/or electron polarons that can provide a transition-metal oxide electrode with the necessary electronic conduction; bad defects include certain ionic defects that block alkali-ion migration, thus hindering the electrochemical performance. A detailed understanding of *defect physics* (i.e., defect thermodynamics and kinetics, defect–defect interaction, etc. and properties governed by defects), including the synthesis–(defect) structure–property relationship, is thus essential to the understanding and design of battery materials.

[*]E-mail: khang.hoang@ndsu.edu

First-principles calculations based on density-functional theory (DFT) [1,2] have been instrumental to the study of defects in solids [3,4] and, particularly, in complex electrode materials [5]. A study of defects in a solid necessarily begins with a careful investigation of the material free of the defects under investigation, which acts as the *host* material for defect calculations (Note that since the host simply acts as a reference, it is not necessarily a perfect bulk compound). This includes the calculation of the atomic and electronic structure, phase stability, and any other bulk properties that may be deemed necessary to understand the physics of the host material. In subsequent calculations, defects are treated within supercell models [6,7], in which a defect is included in a periodically repeated finite volume of the host which itself contains multiple original unit cells.

Systematic first-principles investigations of defects in Li-ion electrode materials were first carried out by Hoang and Johannes [8,9], and the approach has since played an important role in providing a detailed understanding of both widely studied and newly discovered materials and guidelines for designing materials with improved performance. In particular, the approach can predict defect landscapes under different synthesis conditions, provide guidelines for defect characterization and defect-controlled synthesis, uncover the mechanisms for electronic and ionic conduction and for electrochemical extraction and (re-)insertion, and provide a deep understanding of the effects of doping. The term *defect landscape* or *defect energy landscape,* first coined by Hoang and Johannes [8] in 2011, is used to collectively capture the energetics and electronic behavior of all relevant defects that may be present in a host compound, usually manifested in a combined formation-energy plot. It thus implies comprehensive and systematic investigations of all possible point defects that may occur in the material. In such an approach, one carries out the investigations not just because there are defects that need to be studied, but to methodically probe the material at the electronic and atomic level [5].

In the following, we present the methodology for first-principles studies of defects. We then discuss defects that form during materials synthesis through selected examples of defect landscapes as well as guidelines for materials design via defect-controlled synthesis. Next, we discuss defects that are activated during measurements and mechanisms for electronic and ionic conduction in electrode materials. We then regard the electrochemical extraction process in a cathode material during battery use as the creation of electrochemically activated defects and discuss the delithiation mechanism and the derivation of the extraction voltage. Next, we illustrate the effects of doping on the electronic and ionic conduction and the delithiation mechanism and voltage profile. Finally, we end the chapter with some concluding remarks.

## 2 Methodology

First-principles calculations for defects in solids have been widely discussed in the literature. We refer readers to authoritative review articles and book chapters [3,4,10] which discuss in depth the general formalism and various practical aspects. For discussions specific to complex energy materials, see Ref. [5]. In the following, we only highlight aspects most relevant to the discussions in this chapter.

A defect is characterized by its formation energy. In the supercell approach, the formation energy of a defect X in effective charge state $q$ (i.e., with respect to the host lattice) is defined as [3,4]

$$E^f(\mathrm{X}^q) = E_{\mathrm{tot}}(\mathrm{X}^q) - E_{\mathrm{tot}}(\mathrm{host}) - \sum_i n_i \mu_i^* + q(E_{\mathrm{v}} + \mu_{\mathrm{e}}) + \Delta^q, (1)$$

where $E_{\mathrm{tot}}(\mathrm{X}^q)$ and $E_{\mathrm{tot}}(\mathrm{host})$ are the total energies of the defect-containing and defect-free supercells, respectively. $n_i$ is the number of atoms of species $i$ that have been added ($n_i > 0$) or removed ($n_i < 0$) to form the defect. $\mu_i^*$ is the atomic chemical potential, representing the energy of the reservoir with which atoms are being exchanged, and thus dependent on specific experimental conditions under which the defect is formed. $\mu_e$ is the chemical potential of electrons, i.e., the Fermi level, representing the energy of the electron reservoir, and, as a convention, referenced to the valence-band maximum (VBM) in the bulk ($E_{\mathrm{v}}$). Finally, $\Delta^q$ is the correction term to align the electrostatic potentials of the defect-free and defect-containing supercells and to account for finite-size effects on the total energy of charged defects [11].

Under thermodynamic *equilibrium*, the formation energy of a defect directly determines the concentration [3]:

$$c = N_{\mathrm{sites}} N_{\mathrm{config}} \exp\left(\frac{-E^f}{k_{\mathrm{B}}T}\right), \quad (2)$$

where $N_{\mathrm{sites}}$ is the number of high-symmetry sites in the lattice (per unit volume) on which the defect can be incorporated, $N_{\mathrm{config}}$ is the number of equivalent configurations (per site), and $k_{\mathrm{B}}$ is the Boltzmann constant. Equation (2) indicates that, at a given temperature, a defect that has a lower formation energy is more likely to form and occur with a higher concentration. Note that when a material is prepared under *non-equilibrium* conditions, excess defects can be frozen in, and the equilibrium concentration determined via Eq. (2) is only the lower bound.

While the Fermi level $\mu_e$ in Eq. (1) can be treated as a variable, it is not a free parameter. The actual Fermi-level position of the material is determined by the charge-neutrality condition [3]:

$$\sum_i c_i q_i - n_e + n_h = 0, \quad (3)$$

where $c_i$ and $q_i$ are the concentration and charge, respectively, of defect $\mathrm{X}_i$; $n_e$ and $n_h$ are free electron and hole concentrations, respectively; and the summation is over all possible defects present in the material.

Let us illustrate the above general formalism with a specific example, namely a Li vacancy ($V_{\mathrm{Li}}$) in charge state $q$ in the layered oxide cathode material $LiCoO_2$ [see Fig. 1(a)]. From Eq. (1), the formation energy of the vacancy is written as

$$E^f(V_{\mathrm{Li}}^q) = E_{\mathrm{tot}}(V_{\mathrm{Li}}^q) - E_{\mathrm{tot}}(\mathrm{host}) + \mu_{\mathrm{Li}}^* + q(E_{\mathrm{V}} + \mu_e) + \Delta^q, \qquad (4)$$

where $\mu_{\mathrm{Li}}^* = E_{\mathrm{tot}}(\mathrm{Li}) + \mu_{\mathrm{Li}}$, with $E_{\mathrm{tot}}(\mathrm{Li})$ being the total energy per atom of metallic Li. Here, $\mu_{\mathrm{Li}} \leq 0$ eV, with $\mu_{\mathrm{Li}} = 0$ eV meaning that $LiCoO_2$ is in equilibrium with metallic Li.

All the quantities in Eq. (4) can be obtained directly from first-principles total-energy electronic structure calculations, except $\mu_{\mathrm{Li}}$ (see below). For complex transition-metal oxides, calculations based on the Hubbard-corrected DFT+*U* method (an extension of DFT) [12] or a hybrid DFT/Hartree-Fock approach [13] are often employed since it is well known that standard DFT calculations at the local-density (LDA) or generalized gradient (GGA) approximation level [14,15] fails to properly describe the physics of localized electron systems. The hybrid functional approach is computationally more demanding, but it is usually preferred because it treats *all* orbitals in the material on equal footing, unlike DFT+*U* where *a priori* knowledge of the Hubbard *U* value for each orbital in each element in each local chemical environment is required. The DFT/Hartree-Fock mixing parameter and the screening length can be adjusted to match certain calculated properties (e.g., voltage or band gap) to experimental values.

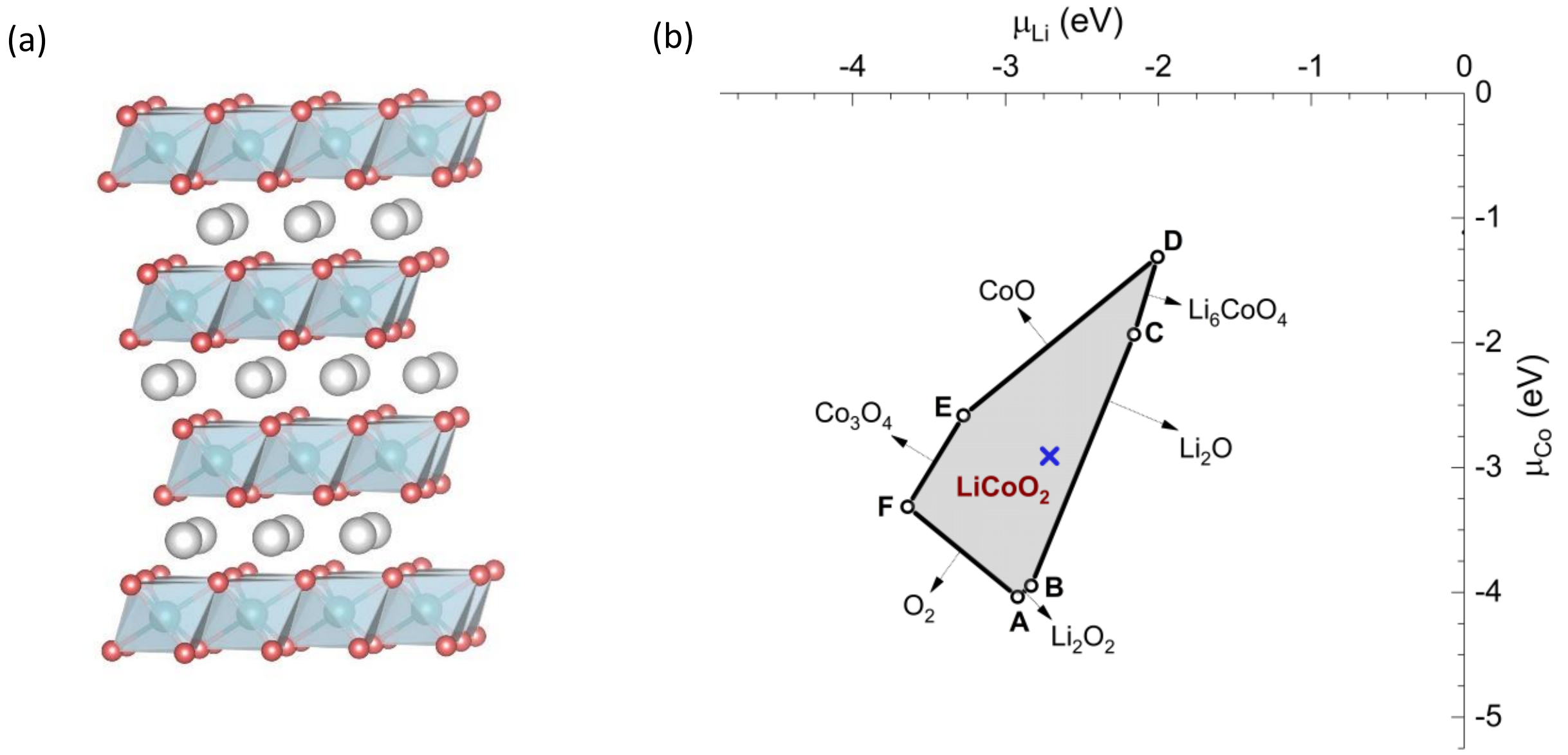

**Figure 1**: (a) Atomic structure of $LiCoO_2$ (trigonal $R\bar{3}m$ space group). Large (gray) spheres are Li, medium (blue) are Co, and small (red) are O. (b) Chemical-potential phase diagram for $LiCoO_2$. Only Li–Co–O phases that define the stability region of $LiCoO_2$, shown as a shaded polygon, are indicated. The diagram is reproduced with data from Ref. [16].

In the above example, $\mu_{Li}$ cannot be calculated without having knowledge of the Li reservoir. Assuming that we are considering defect formation during *synthesis* where Li, Co, and O are in exchange with their respective reservoirs, the upper and lower bounds on $\mu_{Li}$ (as well as $\mu_{Co}$ and $\mu_{O}$) can be determined by requiring that $LiCoO_2$ is stable against competing Li–Co–O phases [16]. Figure 1(b) shows the allowed range of $\mu_{Li}$ and $\mu_{Co}$ values, bound in a polygon in the two-dimensional ($\mu_{Li}$, $\mu_{Co}$) space. For a given point in the polygon, the remaining $\mu_{O}$ variable is determined via the stability condition for the host compound [16]:

$$\mu_{\text{Li}} + \mu_{\text{Co}} + \mu_{\text{O}} = \Delta H(\text{LiCoO}_2), \quad (5)$$

where $\Delta H$ is the formation enthalpy calculated from DFT-based total energies.

Each point in the phase diagram thus corresponds to a set of $\mu_{Li}$, $\mu_{Co}$, and $\mu_{O}$ values which can be employed to represent experimental conditions during synthesis. The actual experimental conditions (if known), on the other hand, can be used to pinpoint the exact location in the phase diagram, which then allows for the calculation of the defect formation energy under those specific conditions. In practice, the exact conditions are often unknown, or the information is incomplete, e.g., that regarding the relative abundance of the constituent elements in the synthesis environment. We usually use information that may be available from experiments such as temperature, the oxygen partial pressure, and/or the presence of certain impurity phases to determine experimentally relevant sets of the atomic chemical potentials. For example, if $Li_2O$ is found as an impurity phase in the preparation of $LiCoO_2$, the synthesis conditions must be very close to the *BC* line in Fig. 1(b). Note that, although such a chemical-potential phase diagram is often constructed based on zero-temperature energies, temperature effects on the formation energy can be included through atomic chemical potentials that involve gaseous phases. For example, $\mu_{O}$ is related to temperature and pressure, and can be set to half of the Gibbs free energy of the $O_2$ gas. The temperature effects for the solids can be ignored, in a first-order approximation, as they are often small [4].

It should be emphasized that the conditions during (e.g., electrical conductivity) *measurements* or during battery *use* can be different from those during synthesis [5]. For example, temperature and the oxygen partial pressure are often significantly different, and not all the constituent elements of $LiCoO_2$ are in exchange with the environment. Care, thus, should be taken when determining the atomic chemical potentials and calculating the formation energy relevant to such more constrained situations. For further discussions and examples, see Ref. [5,17].

## 3 Defects activated during synthesis

### *3.1 Defect landscape and its implications*

Common native defects in electrode materials are small electron and hole polarons, vacancies, interstitials, antisites, and defect complexes. They can be thermally activated during synthesis at high temperatures and get frozen in as the material is cooled down to room temperature, which then act as preexisting/athermal defects in subsequent measurements or use. As a standard procedure, all potential defect configurations are calculated and examined for their structural, electronic, and energetic stability. The stabilization of electron and/or hole polarons, often found in complex oxides, is necessarily determined by the nature of the electronic structure of the host compound at the conduction-band minimum (CBM) and the VBM, respectively, as discussed in detail in Ref. [5] and references therein. The energetics of ionic defects is, in general, strongly dependent on the relative abundance of the host's constituent elements in the synthesis environment (as reflected in the atomic chemical potential values), among other factors.

Figure 2(a) shows the formation energies of low-energy native defects in $LiCoO_2$ [16]. These include small hole polarons ($\eta^{+}$; i.e., low-spin $Co^{4+}$ at the $Co^{3+}$ site), small electron polarons ($\eta^{-}$; i.e., high-spin $Co^{2+}$ at the $Co^{3+}$ site), lithium vacancies ($V_{Li}$) and interstitials ($Li_i$), and lithium ($Li_{Co}$) and cobalt ($Co_{Li}$) antisites. Each ionic defect type has more than one charge states (i.e., *q* values). However, only the following configurations are the true charge states (and hence regarded as elementary defects): $V_{Li}^{-}$ (i.e., a void formed by the removal of a $Li^{+}$ ion), $Li_i^{+}$ (additional $Li^{+}$ ion at an interstitial site); $Li_{Co}^{2-}$ [$Li^{+}$ at the $Co^{3+}$ site; see Fig. 2(b)]; and $Co_{Li}^{+}$ (high-spin $Co^{2+}$ at the $Li^{+}$ site). Other (nominal) charge states are, in fact, complexes consisting of the elementary defects and the small polaron(s). For example, $V_{\mathrm{Li}}^{0}$ is a complex of $V_{\mathrm{Li}}^{-}$ and $\eta^{+}$, $\mathrm{Co}_{\mathrm{Li}}^{0}$ is a complex of $\mathrm{Co}_{\mathrm{Li}}^{+}$ and $\eta^{-}$, and $\mathrm{Li}_{\mathrm{Co}}^{0}$ is a complex of $\mathrm{Li}_{\mathrm{Co}}^{2-}$ and two $\eta^{+}$ [16]. In the absence of electrically active impurities or when they occur with lower concentrations than the charged native defects, the Fermi level of $LiCoO_2$ is at $\mu_e = \mu_e^{\mathrm{int}}$ ("int" is "intrinsic") where charge neutrality is maintained. In the example shown in Fig. 2(a), $\mu_e^{\mathrm{int}}$ is predominantly determined by the dominant (i.e., lowest-energy) charged native defects $\eta^{+}$ and $V_{\mathrm{Li}}^{-}$, and the material can be expressed in terms of the chemical formula $\mathrm{Li}_{1-z}\mathrm{Co}_z^{4+}\mathrm{Co}_{1-z}^{3+}\mathrm{O}_2$.

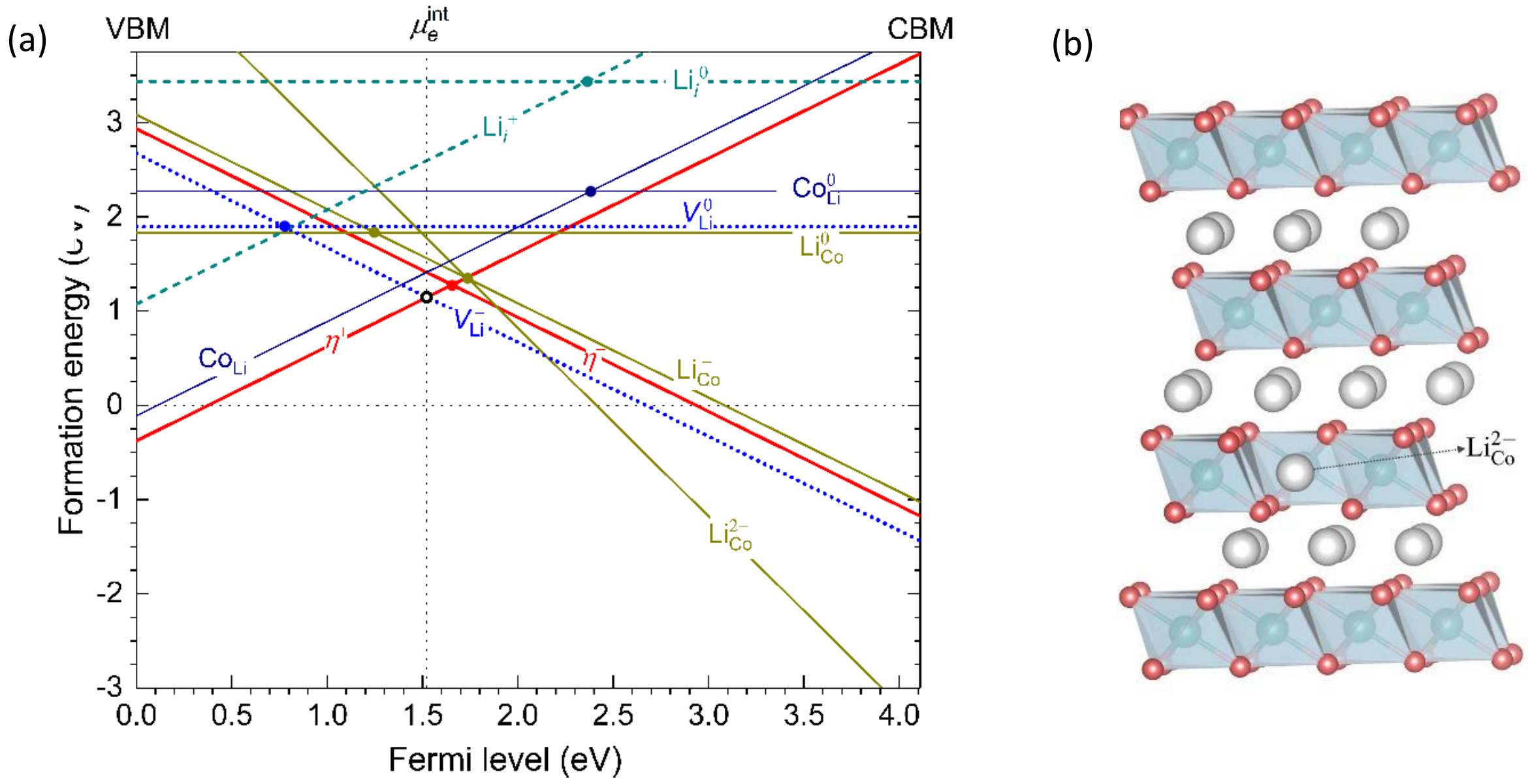

**Figure 2**: (a) Formation energies of low-energy native defects in $LiCoO_2$, plotted as a function of the Fermi level from the VBM to the CBM. The energies are obtained at point $X$ in the phase diagram [indicated by a cross mark in Fig. 1(b)]. The slope indicates charge state $q$: positively (negatively) charged defects have positive (negative) slopes. $\mu_e^{\mathrm{int}}$ is the Fermi-level position determined by the charge neutrality condition involving the native defects. (b) Structure of the Li antisite defect configuration $\mathrm{Li}_{\mathrm{Co}}^{2-}$. Reproduced with data from Ref. [16].

It is clear from the defect landscape for $LiCoO_2$ that a change from one (nominal) defect charge state to another is always associated with polaron formation, and thus the native defects cannot act as sources of band-like electrons and holes. In addition, $\mu_e^{\mathrm{int}}$ is far from both the band edges, i.e., band-like carriers are likely negligible, and some charged native defects have non-negative formation energies only in a small range of the Fermi-level values near midgap. These features indicate that the electronic conduction must occur via hopping of small polarons, and the material cannot be doped n- or p-type like a conventional semiconductor where the Fermi level can be at or very close to the band edges. The Fermi level can be shifted, e.g., via doping, within a certain range, but any attempt to deliberately shift the Fermi level all the way to the VBM or CBM will lead to spontaneous formation of native defects that counteract the effects of shifting [8,9]. The defect landscape thus has important implications for the electrical conduction and doping mechanisms as further discussed in Secs. 4 and 6.

*3.2 Materials design via defect-controlled synthesis*

A defect landscape is a function of the atomic chemical potentials, so it changes when moving from one point to another in the phase diagram. The scenario shown in Fig. 2(a) is thus not the only one that may occur in $LiCoO_2$. As reported in Ref. [16], under the conditions at points *A* and *B*, for example, the lowest-energy elementary native defects in $LiCoO_2$ are $\eta^+$ and $Li_{Co}^{2-}$, whereas they are $Co_{Li}^+$ and $\eta^-$ at points *C* and *D* or $Co_{Li}^+$ and $V_{Li}^-$ at points *E* and *F*. To minimize the formation of cobalt antisites, one thus needs to avoid preparing the material under the conditions in or near the region enclosed by *C*, *D*, *E*, and *F*. To obtain Li-overstoichiometric $LiCoO_2$, one needs to prepare the material under the conditions close to the *BC* line and points *A* and *B*. This is consistent with the fact that the impurity phase $Li_2O$ is often observed when preparing $LiCoO_2$ in Li-excess (Co-deficient) environments [18,19]. Based on the defect landscape obtained at point *B*, for example, the chemical formula for the Li-overstoichiometric samples can be written as $Li_{1+\delta}Co_{1-\delta}O_2$ or, more explicitly, as $Li[Co^{3+}_{1-3\delta}Li^{+}_{\delta}Co^{4+}_{2\delta}]O_2$ where each $Li_{Co}^{2-}$ is charge compensated by two $\eta^+$ (i.e., $Co^{4+}$), which is the neutral complex $Li_{Co}^0$. In all the mentioned defect landscapes for $LiCoO_2$, the lowest-energy defects in the material always contain low-spin $Co^{4+}$ (in the form of $\eta^+$) and/or high-spin $Co^{2+}$ (in the form of $\eta^-$ or $Co_{Li}^+$) [16], which is consistent with the presence of localized magnetic moments observed in experiments [20]. The example here illustrates how one can tune the defect landscape in a material by tuning the synthesis conditions, and how the investigation of different possible defect landscapes provides guidelines for experimental defect characterization.

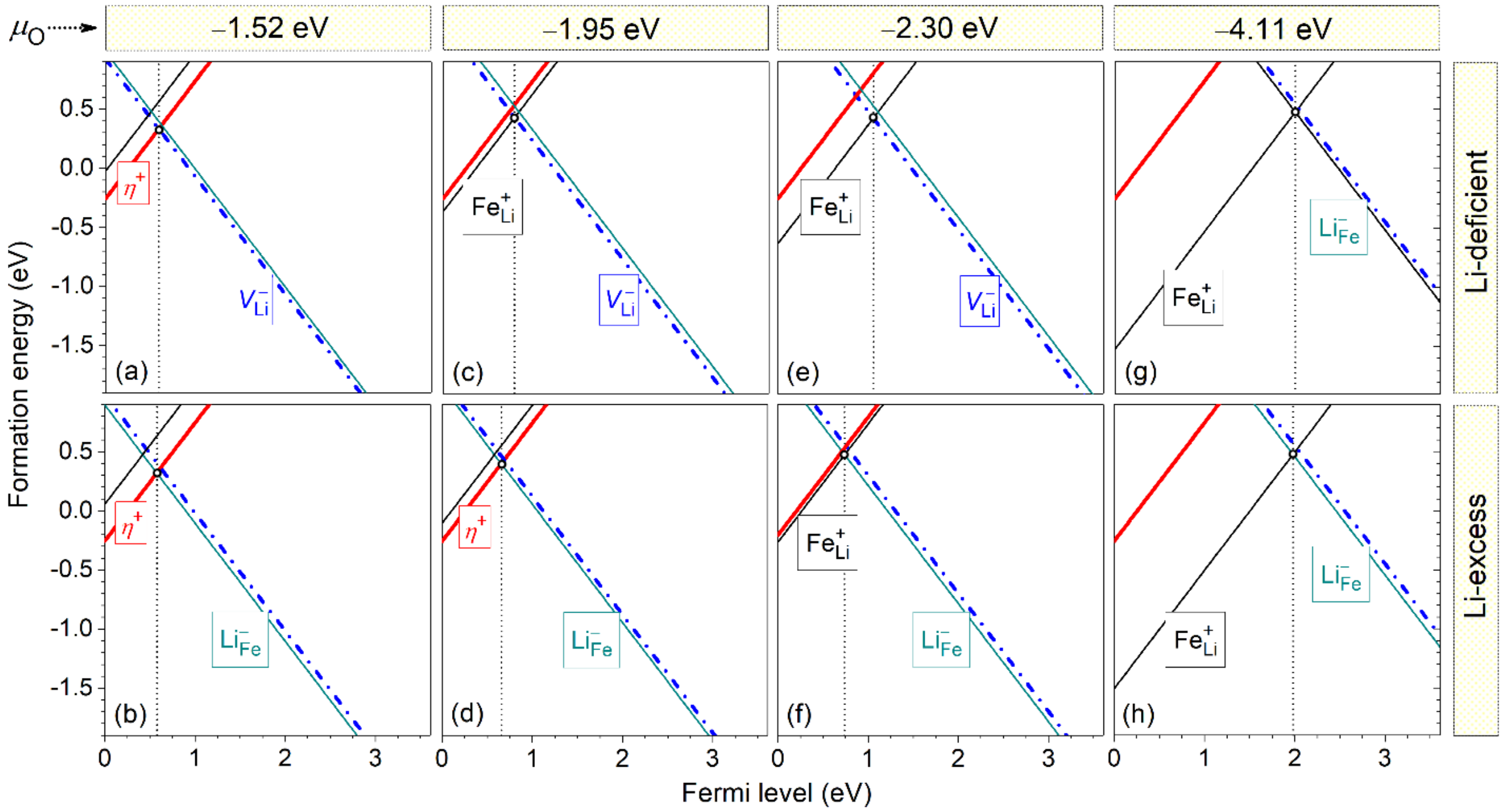

**Figure 3**: Formation energies of low-energy positively and negatively charged defects in $LiFePO_4$, plotted as a function of the Fermi level from the VBM to the CBM. The energies are obtained for different oxygen chemical potential ($\mu_O$) values and in the Li-deficient or Li-excess environment. Reprinted with permission from Ref. [5]; original data reported in Ref. [8].
Figure 3 is another example, reported in Ref. [8], showing how the defect landscape in the olivine phosphate cathode material $LiFePO_4$ changes in going from the lowest to highest possible $\mu_O$ value and from the Li-deficient to Li-excess environment. Lower $\mu_O$ values (more reducing environments) are usually associated with higher temperatures and/or lower oxygen partial pressures and/or the presence of oxygen-reducing agents, whereas higher $\mu_O$ values (more oxidizing environments) are usually associated with lower temperatures and/or higher oxygen partial pressures. It is clear from Fig. 3 that $Fe_{Li}^{+}$ (i.e., $Fe^{2+}$ at the $Li^+$ site) is the dominant positively charged defect in a majority of the defect landscapes, which is consistent with the fact that iron antisites are often observed in $LiFePO_4$ samples [21–24]. These immobile defects are known to block the one-dimensional Li channels in the olivine structure and thus reduce the electrochemical performance—a serious problem in olivine-based Li-ion batteries.

The solution to the above problem is to minimize the formation of iron antisites by tuning the synthesis conditions. Hoang and Johannes [8] in 2011 predicted that $Fe_{Li}^{+}$ has the lowest concentration (i.e., highest formation energy) under the most oxidizing and Li-excess conditions, specifically at/near a point in the Li–Fe–P–O phase diagram where $LiFePO_4$ is in equilibrium with $Li_3Fe_2(PO_4)_3$ and $Li_3PO_4$. Under those conditions, $\eta^+$ (i.e., $Fe^{3+}$ at the $Fe^{2+}$ site) and $Li_{Fe}^{-}$ (i.e., $Li^+$ at the $Fe^{2+}$ site) are the dominant defects, and the concentration of iron antisites is lowest and can be negligible; see Fig. 3(b). The prediction was eventually confirmed by Park et al. [25] in 2016. The Li excess can also open up additional Li diffusion paths perpendicular to the one-dimensional Li channels in the material [25].

## 4 Defects activated during measurements

As discussed in Ref. [5] and references therein, electronic conduction in Li-ion battery electrode materials often occurs via hopping of small polarons; and ionic conduction often occurs via migration of negatively charged Li vacancies and/or positively charged Li interstitials. These current-carrying defects can be *preexisting/athermal* in the material or *thermally activated* during electrical conductivity measurements. Athermal defects may include native defects that occur during synthesis and get trapped in materials (as discussed above), including those that act as charge-compensation centers in materials doped with electrically active impurities (see Sec. 6), and electrochemically activated (positively charged) hole polarons and (negatively charged) Li vacancies present in partially delithiated compounds (see Sec. 5).

The electronic (ionic) conductivity associated with a current-carrying electronic (ionic) defect can be defined as

$$\sigma = qmc, \quad (6)$$

where $q$, $m$, and $c$ are the defect's charge, mobility, and concentration, respectively. The mobility can be assumed to be thermally activated, i.e.,

$$m = \frac{m_0}{T} \exp\left(-\frac{E_m}{k_B T}\right), \quad (7)$$

where $m_0$ is a prefactor and $E_m$ is the defect migration barrier. The concentration $c$ can include both thermally activated and athermal defects [16,26],

$$c = c_a + c_t = c_a + c_0 \exp\left(-\frac{E^f}{k_B T}\right), \qquad (8)$$

where $c_a$ is the athermal concentration consisting of defects that preexist in the material before the conductivity measurements, $c_t$ is the concentration of defects that are thermally activated during the measurements at finite temperatures, $c_0$ is a prefactor, and $E^f$ is the defect formation energy.

The above expressions indicate that when the athermal defects are dominant, i.e., $c_a \gg c_t$, the activation energy of the conductivity includes only the migration barrier, i.e.,

$$E_a = E_m. \quad (9)$$

This corresponds to the *extrinsic* (low-temperature) region often shown in Arrhenius plots of $\ln(\sigma T)$ versus $1/T$.

When the thermally activated defects are dominant, i.e., $c_t \gg c_a$, the activation energy includes both the formation energy and migration barrier, i.e.,

$$E_a = E^f + E_m, \quad (10)$$

which corresponds to the *intrinsic* (high-temperature) region in Arrhenius plots. The intrinsic and extrinsic regions are joined by a convex knee [27]. See further discussions in Ref. [5].

The migration barrier of the ionic current-carrying defects (e.g., Li vacancies and/or interstitials) can be computed using standard techniques such as the nudged elastic band method [28]. In battery materials, Li migration often occurs via a (mono)vacancy mechanism, particularly via $V_{\mathrm{Li}}^-$, at least at low vacancy concentrations. The movement of $V_{\mathrm{Li}}^-$ in one direction is equivalent to that of a $\mathrm{Li}^+$ ion in the opposite direction. In layered oxides, Li migration can also occur via a divacancy mechanism in which the movement of one $V_{\mathrm{Li}}^-$ happens in the presence of another $V_{\mathrm{Li}}^-$ [16,29] The divacancy mechanism is expected to be dominant in partially delithiated layered oxides where the vacancy concentration is high [16,30,31]. The migration of a small polaron between two positions $q_A$ and $q_B$, on the other hand, can be described by the transfer of its lattice distortion and the migration barrier is obtained by computing the total energies of a set of supercell configurations linearly interpolated between $q_A$ and $q_B$ and identifying the energy maximum [32–34].

The defect formation energy can be calculated following the methodology described in Sec. 2 *but* with a set of atomic chemical potentials associated with the experimental conditions under which the measurements are carried out. It is, however, often difficult to assess the exact conditions, including the defect formation mechanism, in a measurement. As a result, the formation energy part in Eq. (10) is usually calculated explicitly only in cases where it is independent of the atomic chemical potentials and the Fermi-level position, e.g., when the thermally activated, current-carrying defects are known to be created via a Frenkel or full-Schottky defect mechanism or via an electron–hole polaron pair mechanism [5].

From expressions (6)–(8), it is clear that the electrical conductivity in electrode materials can be increased by one or more of the following: (i) increasing the athermal current-carrying defects (e.g., via defect-controlled synthesis, doping, and/or electrochemical extraction of some Li from the material beforehand); (ii) increasing the thermally activated defects (e.g., by lowering the formation energy of the current-carrying defects via doping with suitable electrically active impurities that shift the Fermi level in the right direction; see Sec. 6); and (iii) lowering the defect migration barrier (in this case, doping might or might not help). Note that thermally activated defects may matter less to battery use at room temperature as the electronic and ionic conduction in the cathode can rely on available athermal defects [e.g., $\eta^+$ and $V_{\mathrm{Li}}^-$ that are activated during a (pre-)delithiation step; see below], but they become more important in materials used in higher-temperature applications such as solid-oxide fuel cells [17].

## 5 Defects activated during battery use

### *5.1 Lithium extraction mechanism*

In Li-ion batteries, the Li extraction/(re-)insertion processes during charge/discharge can be regarded as corresponding to the creation of electrochemically activated defects in the electrode materials [30]. Let us take the delithiation reaction in a $LiMnO_2$ cathode as an example:

$$LiMnO_2 \rightarrow Li_{1-x}MnO_2 + xLi^+ + xe^-. \quad (11)$$

Here, the extracted $Li^+$ ions dissolve into the electrolyte and the electrons move in the opposite direction to the outer circuit. The extraction of lithium ($Li^+$ plus $e^-$) from the electrode corresponds to the electrochemical activation of Li vacancies ($V_{\mathrm{Li}}^0$) in $LiMnO_2$.

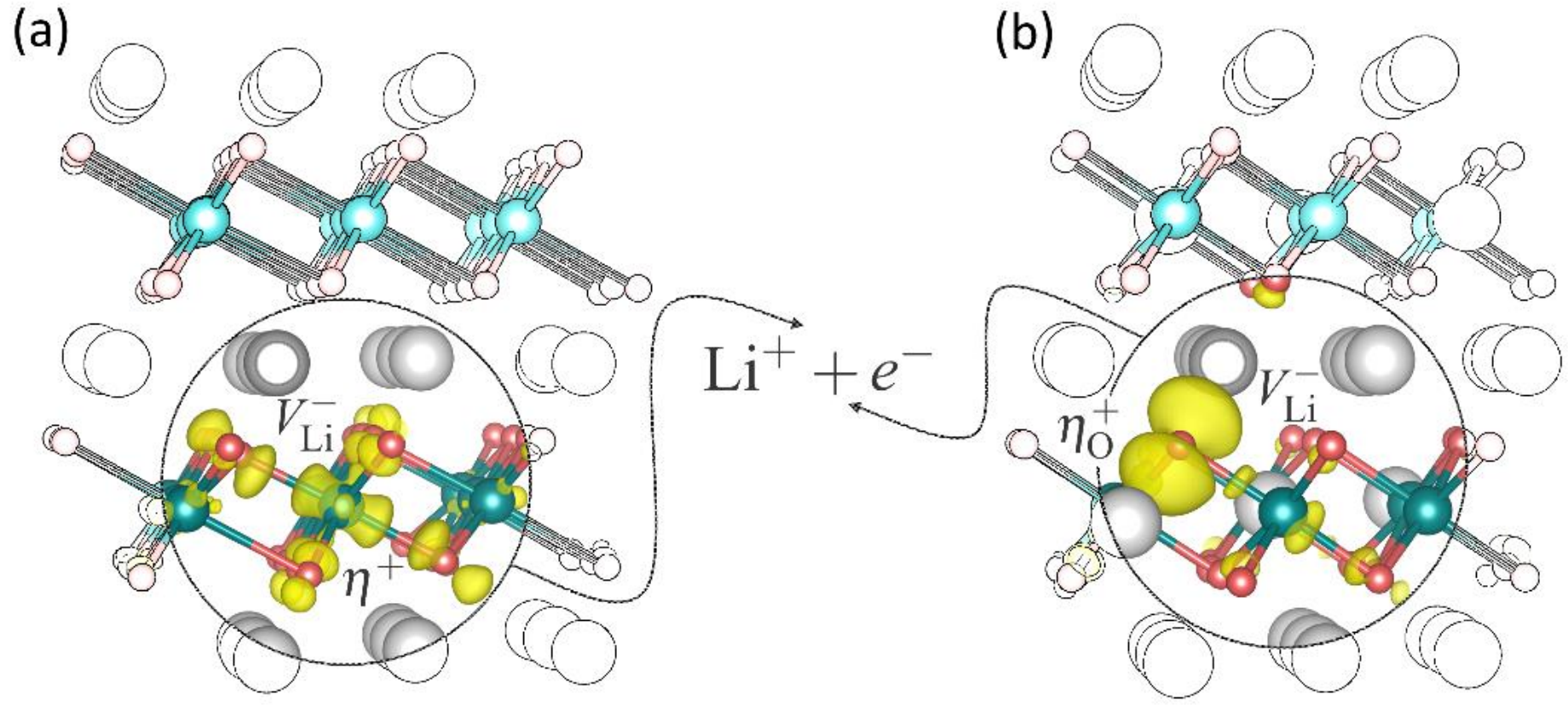

**Figure 4**: Mechanisms for Li extraction involving oxidation (a) at the transition-metal site in $LiMnO_2$ and (b) at the oxygen site in $Li_2MnO_3$ [30]. Large (gray) spheres are Li, medium (blue) are Mn, and small (red) are O. Charge densities associated with the hole polarons $\eta^+$ (i.e., $Mn^{4+}$) and $\eta_O^+$ (i.e., $O^-$) are visualized as (yellow) isosurfaces. Reprinted with permission from [5].

Figure 4 shows the structure of $V_{\mathrm{Li}}^0$ in the layered oxide $LiMnO_2$ and that in the Li-rich layered oxide $Li_2MnO_3$ and their implications on the delithiation mechanism [30]. Most alkali-ion battery cathode materials [8,16,31,35–38] are like $LiMnO_2$ according to which $V_{\rm Li}^{0}$ is a complex of $V_{\rm Li}^{-}$ and $\eta^{+}$ (a hole polaron at the transition-metal site). In these materials, oxidation thus occurs on the transition-metal ion and the removal of a $Li^+$ ion (i.e., the creation of $V_{\rm Li}^{-}$) is charge compensated by the formation of $\eta^{+}$. In $Li_2MnO_3$, on the other hand, $V_{\rm Li}^{0}$ is a complex of $V_{\rm Li}^{-}$ and $\eta_{\rm O}^{+}$ with the latter being an oxygen-hole polaron (i.e., $O^-$) *bound* to the former. $\eta_{\rm O}^{+}$ is stable only in the presence of a nearby $V_{\rm Li}^{-}$ or some other negatively charged defect. In this Li-rich material, where oxygen is undercoordinated compared to that in $LiMnO_2$, the intrinsic delithiation mechanism thus involves oxidation at the oxygen site [30]. $Li_2MnO_3$ has been the prototypical material for the study of anionic redox. The different mechanisms in $LiMnO_2$ and $Li_2MnO_3$ can be traced back to the difference in their electronic structure; see further discussions in [30]. Note that, at high vacancy concentrations (i.e., $Li_{2-x}MnO_3$ with high $x$ values), vacancy–vacancy interactions might affect the detailed structure of the vacancies and hence the delithiation mechanism. A physics similar to that of $Li_2MnO_3$ has also been found in other battery materials with anionic redox such as $Li_8ZrO_6$ and $Na_2ZrO_3$ [39,40].

### *5.2 Lithium extraction voltage*

The methodology presented in Sec. 2 and the connection between the defect configuration $V_{\rm Li}^{0}$ and the delithiation mechanism also provide an intuitive way to derive the extraction voltage based on the energetics of $V_{\rm Li}^{0}$, as described in Ref. [30].

From Eq. (4), an expression for the formation energy of arbitrarily $x$ neutral lithium vacancies becomes

$$E^f(xV_{\rm Li}^{0}) = E_{\rm tot}(xV_{\rm Li}^{0}) - E_{\rm tot}({\rm host}) + x\mu_{\rm Li}^{*}. \quad (12)$$

Note, again, that the creation of $V_{\rm Li}^{0}$ in the cathode material is equivalent to the removal of Li (which, in this case, a complex of $Li^+$ and an electron) from the material. Assuming the cathode (host) material is in equilibrium with a $Li/Li^+$ anode (which acts as the reservoir of $Li^+$ ions) and the external power source with a voltage $V$ (which acts as the reservoir of electrons), the Li chemical potential can be expressed as

$$\mu_{\rm Li}^{*} = E_{\rm tot}({\rm Li}) - eV. \quad (13)$$

During the delithiation process, the Li vacancies can be assumed to readily form under the influence of the extraction voltage, i.e.,

$$E^f(xV_{\rm Li}^{0}) = 0. \quad (14)$$

From the above expressions, the lithium-extraction voltage can be expressed in terms of the total energies as [30]

$$V = \frac{E_{\rm tot}(xV_{\rm Li}^{0}) - E_{\rm tot}({\rm host}) + xE_{\rm tot}({\rm Li})}{xe}. \quad (15)$$

Here, $x$ can take any value to describe the lithium-content difference between any two intercalation limits, and the host material can be any starting chemical composition. In that case, $V$ becomes the average voltage between the two limits. The above expression, derived by Hoang [30], is equivalent to that derived by Aydinol et al. [41] using a different approach.

From the general expression (15), one can write a specific formula for *any* specific material and supercell model. The voltage for $LiCoO_2$, for example, can be written as

$$V = \frac{E_{\text{tot}}(\text{Li}_{1-x}\text{CoO}_2) - E_{\text{tot}}(\text{LiCoO}_2) + xE_{\text{tot}}(\text{Li})}{xe}. \quad (16)$$

The above-mentioned discovery allows us to uncover the delithiation mechanism and voltage in a cathode material just by examining the $V_{\text{Li}}^0$ defect configuration. Modifying the structure and energetics of $V_{\text{Li}}^0$ would thus lead to a modification of the delithiation mechanism and voltage profile. Note that the mechanism for Li re-insertion can be studied in a similar way by examining the structure of lithium interstitials ($\text{Li}_i^0$) in the fully or partially delithiated compound which now acts the host material for the re-insertion process [30].

## 6 Materials design via doping

Doping is the intentional introduction of impurities into a material, usually during synthesis. It has been widely employed to optimize the performance of battery materials. In a first-principles study of doping, the first task after the investigations of the bulk properties and native defects is to determine the *lattice site preference* of an impurity (i.e., dopant) in the host lattice [9,42,43]. That is to find which lattice site, among all possible sites at which a dopant can be incorporated, is energetically most favorable when the doped material is prepared under certain conditions.

Comprehensive first-principles studies [9,42,43] show that the lattice site preference depends strongly on the relative abundance of the host's constituent elements in the synthesis environment, and not just the ionic-radius difference between the dopant and the substituted host atom. For transition-metal impurities, the lattice site preference is also determined by the dopant's charge and spin states which are strongly coupled to the local lattice environment and can be strongly affected by the presence of co-dopant(s).

### *6.1 Tuning the electronic and ionic conduction*

In electrode materials, doping can significantly change the concentration of current-carrying native defects and hence the electronic and ionic conductivities. As discussed in [5] and references therein, there are two effects on the defect concentration. One is the charge-compensation effect that occurs during synthesis. Impurities when introduced into the material can be positively or negatively charged, and native defects with the opposite charge are formed to maintain charge neutrality. Both the impurities and the charge-compensating native defects are present in the material and act as *preexisting/athermal* defects in subsequent measurements or use. The other effect involves the shift of the Fermi level as the charge neutrality condition is re-established in the presence of the electrically active impurities. This leads to an increase or decrease in the formation energy and hence the concentration of *thermally activated* current-carrying defects.

Figure 5 illustrates how the formation energy of $\eta^+$ and $V_{\rm Li}^-$ gets modified in the case of acceptor- or donor-like doping. Here, $\eta^+$ and $V_{\rm Li}^-$ are assumed to be the dominant native defects which determine the Fermi level $\mu_e^{\rm int}$ of the undoped material. When incorporated with a concentration higher than that of $V_{\rm Li}^-$, an acceptor-like impurity will shift the Fermi level from $\mu_e^{\rm int}$ slightly toward the VBM, thus decreasing (increasing) the formation energy of $\eta^+$ ($V_{\rm Li}^-$). For a donor-like impurity, the Fermi level will be shifted slightly toward the CBM, thus decreasing (increasing) the formation energy of $V_{\rm Li}^-$ ($\eta^+$) [5,9]. Such a change in the formation energy of $\eta^+$ ($V_{\rm Li}^-$) will change the activation energy in the intrinsic region and hence the electronic (ionic) conductivity accordingly. The argumentation still holds even when $\eta^+$ and $V_{\rm Li}^-$ are not the dominant native defects. In that case, the dopant concentration needs to be higher than that of the lowest-energy charged native defect.

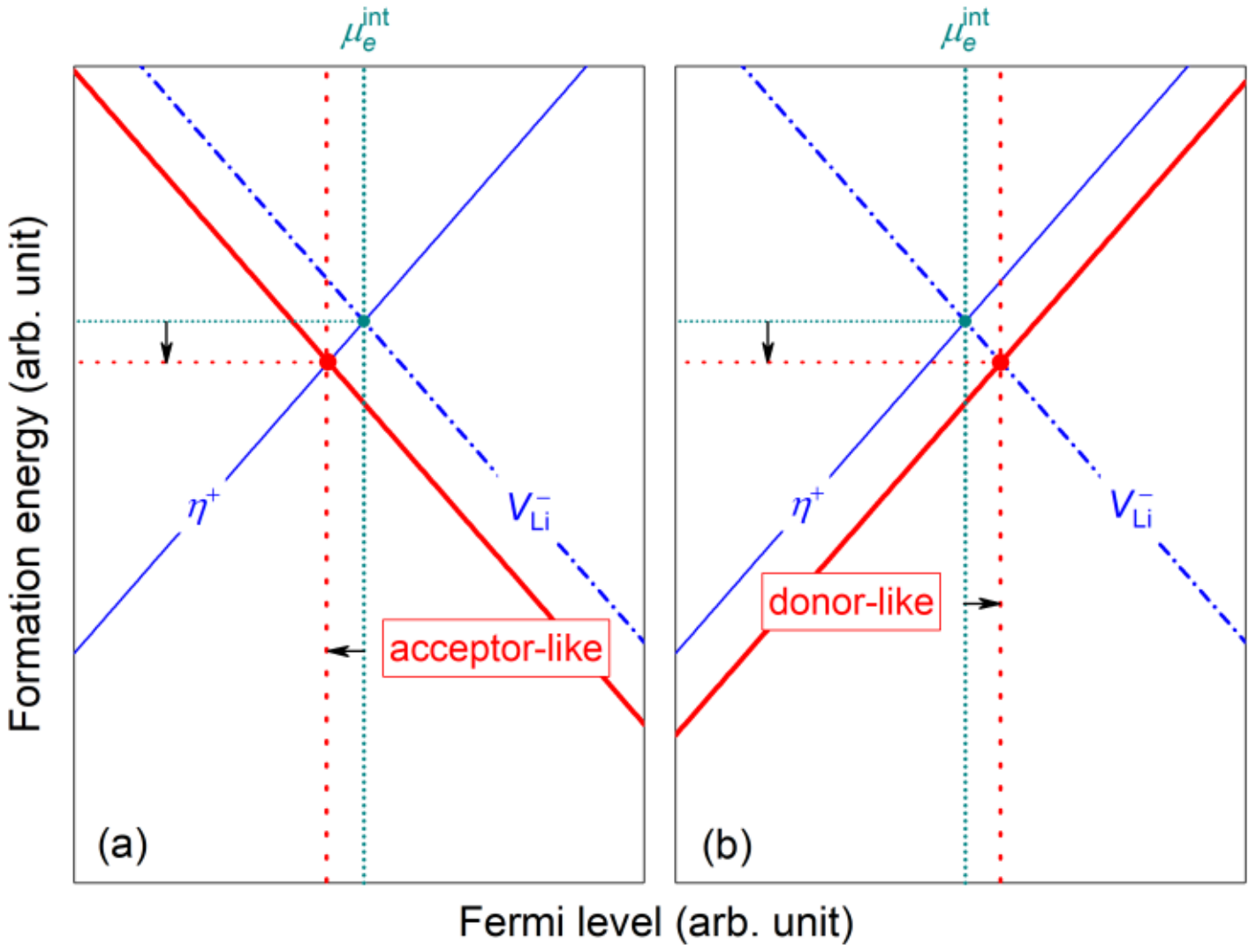

**Figure 5**: Schematic illustration of (a) acceptor-like or (b) donor-like doping and its effects on current-carrying native defects (e.g., $\eta^+$ and $V_{\rm Li}^-$) in an electrode material. Adapted from Ref. [9].

Let us take Mg-doped $LiCoO_2$ as an example. When incorporated into the material at the Co site, Mg is stable as $\rm Mg_{Co}^-$. This acceptor-like impurity can shift the Fermi level slightly toward the VBM, like in Fig. 5(a), and is charge compensated by $\eta^+$ during synthesis [42]. In addition to the increased athermal $\eta^+$ concentration, the Fermi-level shift lowers the formation energy of $\eta^+$ and hence the activation energy for electronic conduction during measurements. Experimentally, doping $LiCoO_2$ with Mg has been shown to enhance the electronic conductivity, and the activation energy decreases from 0.16 eV in $Li_{1.0}CoO_2$ to 0.11 eV in $Li_{1.0}Co_{0.97}Mg_{0.03}O_2$ [44,45]. The activation energy of the Mg-doped sample is very close to the migration barrier of $\eta^+$ (0.10 eV) [16], indicating that $c_a \gg c_t$ and thus $E_a \approx E_m$, as in Eq. (9).

The migration barrier of current-carrying defects can also be tuned via doping. However, if the migration barrier gets reduced (or increased) near a dopant, it does not necessarily mean that the conductivity will get increased (reduced) accordingly. It is important to determine if the effect is local or global, as the defects still need to move across the material, and if there are alternative migration paths. Care should be taken when interpreting the computational results.

## *6.2 Modification of the delithiation mechanism and voltage*

Doping can also significantly modify the delithiation mechanism and voltage. Here, again, we use $Li_2MnO_3$ as an example. Undoped $Li_2MnO_3$ is known to be electrochemically activated only at high voltages [46] and shows very limited electrochemical capacity [47]. As argued in Refs. [30] and [43], the difficulty in activating $Li_2MnO_3$ can be ascribed to its unconventional delithiation mechanism (see Sec. 5), including the high extraction voltage and a lack of percolation pathways for electronic conduction in the bulk at the onset of delithiation.

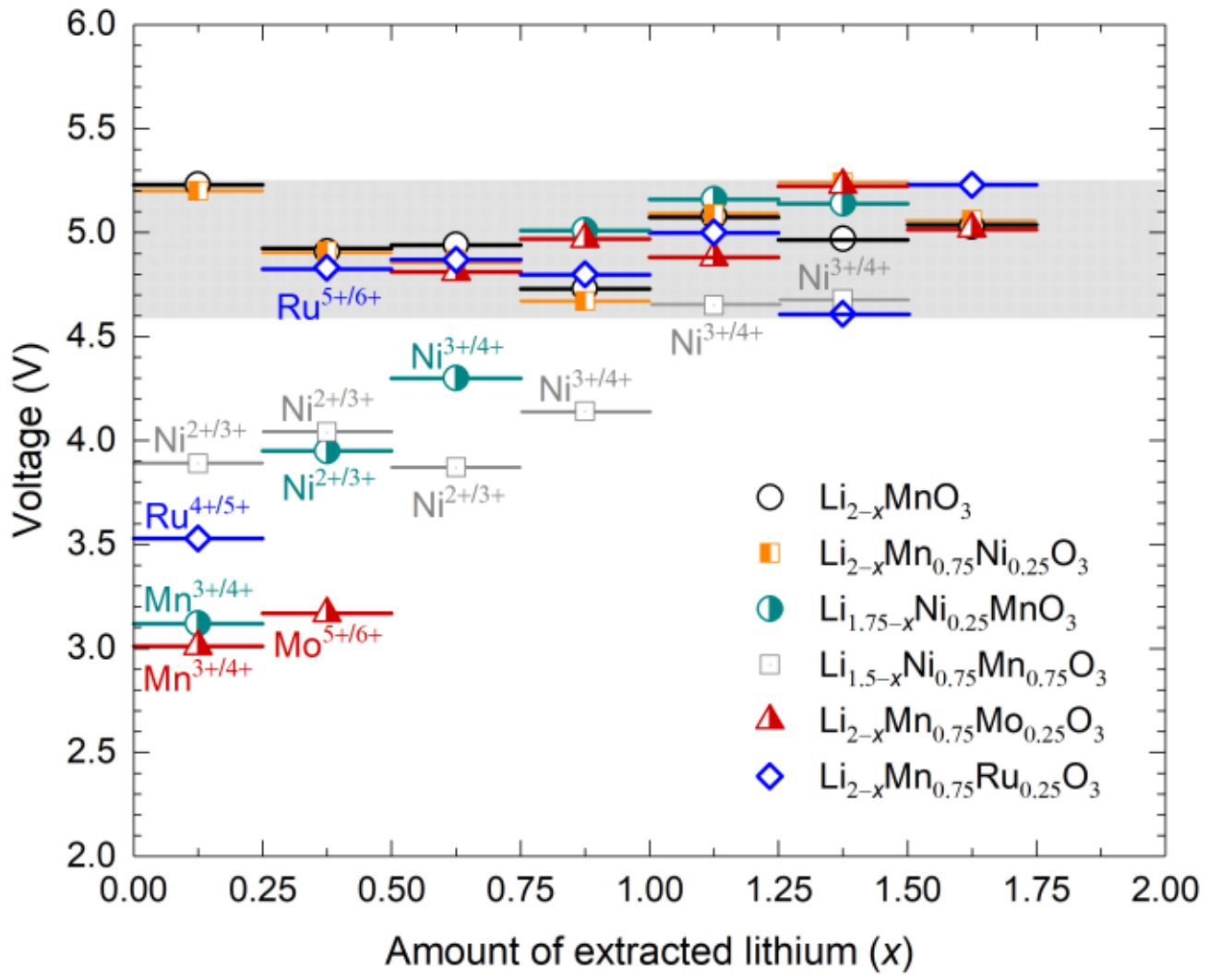

**Figure 6**: Voltage profiles of undoped $Li_2MnO_3$ and $Li_2MnO_3$ doped with Ni, Mo, or Ru. The redox couples associated with different voltage points are indicated; unmarked voltage points in the shaded area are those associated with the redox activity on oxygen. Reprinted with permission from Ref. [43], copyright 2017 by the American Physical Society.

Figure 6 shows the calculated voltage profiles of undoped $Li_2MnO_3$ and $Li_2MnO_3$ heavily doped with Ni, Mo, and Ru [51]. The specific chemical compositions for the calculations are based on the lattice site preference of the impurities determined in lightly doped $Li_2MnO_3$. These fully lithiated compounds include (i) $Li_2Mn_{1-z}Ni_zO_3$ with Ni stable as $Ni^{4+}$ at the Mn site, (ii) $Li_{2-z}Ni_zMnO_3$ with Ni stable as $Ni^{2+}$ at the Li (2$b$) site and charge compensated by $\eta^-$ (i.e., $Mn^{3+}$), (iii) $Li_{2-2z}Ni_{3z}Mn_{1-z}O_3$ or, equivalently, $Li[Ni_yLi_{1/3-2y/3}Mn_{2/3-y/3}]O_2$, with one Ni at the Mn site for every two Ni at the Li (2$b$) site, all stable as $Ni^{2+}$, (iv) $Li_2Mn_{1-z}Mo_zO_3$ with Mo stable as $Mo^{5+}$ at the Mn site and charge compensated by $\eta^-$, and (v) $Li_2Mn_{1-z}Ru_zO_3$ with Ru stable as $Ru^{4+}$ at the Mn site; $z = y/2 = 1/4$ in all cases. The voltage associated with the oxygen oxidation, in the case of $Li_2MnO_3$ and $Li_2Mn_{1-z}Ni_zO_3$, is high as expected. However, in those cases with electrochemically active transition-metal impurities, the voltage is much lower, at least during early stages of delithiation (i.e., small *x* values). In certain cases, the interaction between the impurity and the host compound also turns some inactive $Mn^{4+}$ ions of the host into $Mn^{3+}$ which are then oxidized during lithium extraction [43].

Note that the above effects can also be achieved in O-deficient $Li_2MnO_3$. Although the formation energy of oxygen vacancies is relatively high in the bulk, it can get significantly reduced at/near the surface or interface due to the less constrained lattice environment [30]. A high vacancy concentration can thus be obtained, e.g., in samples with high surface-to-volume ratios. Since a neutral oxygen vacancy ($V_O^0$) is a complex of $V_O^{2+}$ and two $\eta^-$ [30], high oxygen vacancy concentrations mean high concentrations of electrochemically active $Mn^{3+}$ whose presence is crucial during the early stages of the delithiation process in $Li_2MnO_3$.

## 7 Concluding remarks

We have discussed a computational approach in which defect physics is used as a *conceptual and theoretical framework* for the understanding and design of battery materials. We start with the methodology for first-principles studies of defects in complex materials, and then consider defects that are activated in a cathode material during synthesis, during measurements, and during battery use. Through these cases, we illustrate the use of first-principles defect investigations to methodically probe the material at the electronic and atomic level to develop a deep understanding of various defect-related processes and guidelines for materials design. Despite the focus on battery materials in this chapter, the approach is general and applicable to any other functional materials in which defect physics drives the properties of interest.

## Acknowledgements

This work used resources of the Center for Computationally Assisted Science and Technology (CCAST) at North Dakota State University, which were made possible in part by NSF MRI Award No. 2019077.